\documentclass{article}
\usepackage{a4wide}
\makeatletter

\newcommand{\LyX}{L\kern-.1667em\lower.25em\hbox{Y}\kern-.125emX\@}

\makeatother

\begin{document}

\title{Does Anti-Parallel Spin Always Contain More Information?}

\author{Sibasish Ghosh\protect\( ^{a\: }\protect \)\thanks{
res9603@isical.ac.in
}, Anirban Roy\protect\( ^{a\: }\protect \)\thanks{
res9708@isical.ac.in
}, Ujjwal Sen\protect\( ^{b\: }\protect \)\thanks{
dhom@bosemain.boseinst.ernet.in
} }

\maketitle
\( ^{a} \)Physics and Applied Mathematics Unit, Indian Statistical Institute,
203 B. T. Road, Calcutta -700035, India 

\( ^{b} \)Dept. of Physics, Bose Institute, 93/1 A.P.C. Road, Calcutta - 700009,
India

\begin{abstract}
We show that the Bloch vectors lying on any great circle is the largest set
\( S_{L} \) for which the parallel states \( \left| \overrightarrow{n},\overrightarrow{n}\right\rangle  \)
can always be exactly transformed into the anti-parallel states \( \left| \overrightarrow{n},-\overrightarrow{n}\right\rangle  \).
Thus more information about \( \overrightarrow{n} \) is \emph{not} extractable
from \( \left| \overrightarrow{n},-\overrightarrow{n}\right\rangle  \) than
from \( \left| \overrightarrow{n},\overrightarrow{n}\right\rangle  \) by any
measuring strategy, for \( \overrightarrow{n}\in S_{L} \). Surprisingly, the
largest set of Bloch vectors for which the corresponding qubits can be flipped
is again \( S_{L} \). We then show that probabilistic exact parallel to anti-parallel
transformation is not possible if the corresponding anti-parallels span the
whole Hilbert space of the two qubits. These considerations allow us to generalise
a conjecture of Gisin and Popescu (Phys. Rev. Lett. 83 432 (1999)).
\end{abstract}
Recently, Gisin and Popescu {[}1{]} revealed that for qubits, there exists a
measuring strategy (see {[}2{]}, {[}3{]}) on the anti-parallel state \( \left| \overrightarrow{n},-\overrightarrow{n}\right\rangle  \)
that extracts more information about an arbitrary Bloch vector \( \overrightarrow{n} \)
than that can be extracted from the parallel state \( \left| \overrightarrow{n},\overrightarrow{n}\right\rangle  \)
{[}4{]}. 

In this paper, we ask whether there exists any proper subset \( S \) of the
unit ball \( B_{3}=\{\overrightarrow{n}:\overrightarrow{n}\in R^{3},\left| \overrightarrow{n}\right| =1\} \)
for which an anti-parallel state selected at random from the set \( A_{S}=\{\left| \overrightarrow{n},-\overrightarrow{n}\right\rangle :\overrightarrow{n}\in S\} \)
carry more information about the Bloch vector \( \overrightarrow{n} \) than
a parallel state from the set \( P_{S}=\{\left| \overrightarrow{n},\overrightarrow{n}\right\rangle :\overrightarrow{n}\in S\} \).
We provide a partial answer to this query. Specifically, we find the largest
set \( S_{L} \) for which there exists a unitary operator \( U=U(S_{L}) \)
such that 
\begin{equation}
\label{1}
U\left| \overrightarrow{n},\overrightarrow{n}\right\rangle \left| M\right\rangle =e^{i\theta (\overrightarrow{n})}\left| \overrightarrow{n},-\overrightarrow{n}\right\rangle \left| N(\overrightarrow{n})\right\rangle 
\end{equation}

for all \( \overrightarrow{n}\in S_{L} \), where \( \left| M\right\rangle  \)
and \( \left| N(\overrightarrow{n})\right\rangle  \) are the states of a possible
ancilla. \( \left| N(\overrightarrow{n})\right\rangle  \) must be independent
of \( \overrightarrow{n} \) except possibly in a phase to satisfy the unitarity
of \( U \) {[}5{]}. Consequently an anti-parallel state chosen at random from
\( A_{S_{L}} \) cannot contain more information about \( \overrightarrow{n} \)
than in the corresponding parallel state from \( P_{S_{L}} \). 

A related query is how far we can go by just flipping the second state in order
to transform \( \left| \overrightarrow{n},\overrightarrow{n}\right\rangle  \)
to \( \left| \overrightarrow{n},-\overrightarrow{n}\right\rangle  \). Here
one may think that the largest set of Bloch vectors would in this case be a
very small subset of \( S_{L} \). Surprisingly, as we would show here, this
conjecture is not true: the largest set is again \( S_{L} \). 

We shall show here that universal exact machines for the transformations \( \left| \overrightarrow{n},\overrightarrow{n}\right\rangle  \)
to \( \left| \overrightarrow{n},-\overrightarrow{n}\right\rangle  \) and \( \left| \overrightarrow{n},-\overrightarrow{n}\right\rangle  \)
to \( \left| \overrightarrow{n},\overrightarrow{n}\right\rangle  \) do not
exist. Therefore, in the same vein as one considered deterministic inexact {[}6{]}
and probabilistic exact cloning {[}7,8{]} when faced with the no-cloning theorem
{[}9{]}, one can consider deterministic inexact and probabilistic exact parallel
to anti-parallel (and antiparallel to parallel) machines. Gisin {[}10{]} has
considered the deterministic inexact case. In this paper we consider the probabilistic
exact machines for parallel to anti-parallel transformations. The motivation
behind such consideration is to provide basis for a conjecture on the type of
sets of Bloch vectors \( \overrightarrow{n} \) for which the anti-parallel
state \( \left| \overrightarrow{n},-\overrightarrow{n}\right\rangle  \) contains
more information about \( \overrightarrow{n} \) than the corresponding parallel
state \( \left| \overrightarrow{n},\overrightarrow{n}\right\rangle  \). We
also provide ground for a conjecture made by Gisin and Popescu in {[}1{]}. 

Let us first demonstrate the non-existence of a machine that transforms \( \left| \overrightarrow{n},\overrightarrow{n}\right\rangle  \)
to \( \left| \overrightarrow{n},-\overrightarrow{n}\right\rangle  \) universally.
Suppose that there exists a unitary operator \( U^{\prime } \) such that 
\begin{equation}
\label{2}
U^{\prime }\left| \overrightarrow{n},\overrightarrow{n}\right\rangle \left| M\right\rangle =e^{i\vartheta (\overrightarrow{n})}\left| \overrightarrow{n},-\overrightarrow{n}\right\rangle \left| N\right\rangle 
\end{equation}

for all \( \overrightarrow{n} \), where \( \left| N\right\rangle  \) is independent
of \( \overrightarrow{n} \). First, let us display the action of \( U^{\prime } \)
on the (orthogonal) basis \( \left| 00\right\rangle  \), \( \left| 11\right\rangle  \)
and \( \frac{1}{\sqrt{2}}(\left| 01\right\rangle +\left| 10\right\rangle ) \)
of the linear space spanned by the parallel states:

\[
U^{\prime }\left| 00\right\rangle \left| M\right\rangle =e^{ia}\left| 01\right\rangle \left| N\right\rangle \]
\begin{equation}
\label{3}
U^{\prime }\left| 11\right\rangle \left| M\right\rangle =e^{ib}\left| 10\right\rangle \left| N\right\rangle 
\end{equation}
\[
U^{\prime }\frac{1}{\sqrt{2}}(\left| 01\right\rangle +\left| 10\right\rangle )\left| M\right\rangle =e^{ic}(c_{1}\left| 00\right\rangle +c_{2}\left| 11\right\rangle )\left| N\right\rangle \]

For an arbitrary parallel state \( \left| \overrightarrow{n},\overrightarrow{n}\right\rangle \: (\left| \overrightarrow{n}\right\rangle =e^{i\alpha _{n}}(\cos \frac{\theta _{n}}{2}\left| 0\right\rangle +e^{i\phi _{n}}\sin \frac{\theta _{n}}{2}\left| 1\right\rangle ),\: 0\leq \theta _{n}\leq \pi ,\: 0\leq \phi _{n}<2\pi ) \),
checking for unitarity and linearity, it is easy to show that condition (2)
will be satisfied only for those \( \overrightarrow{n} \)'s for which \( \phi _{n}=\frac{a-b+\pi }{2}+l\pi \: (l=0,\pm 1,\pm 2,....) \). 

Thus we have shown that there exists a machine which (unitarily) transforms
\( \left| \overrightarrow{n},\overrightarrow{n}\right\rangle  \) to \( \left| \overrightarrow{n},-\overrightarrow{n}\right\rangle  \)
for all \( \overrightarrow{n} \) lying on any great circle of the Bloch sphere.
But is this the largest set of Bloch vectors \( \overrightarrow{n} \) for which
the transformation \( \left| \overrightarrow{n},\overrightarrow{n}\right\rangle  \)
to \( \left| \overrightarrow{n},-\overrightarrow{n}\right\rangle  \) is possible
using a single unitary operator? One would be reluctant to believe it as we
have constrained \( U^{\prime } \) to transform at least two orthogonal states
\( \left| 00\right\rangle  \) and \( \left| 11\right\rangle  \) to their corresponding
anti-parallels. But we shall show that a great circle is the largest set of
Bloch vectors for which the corresponding parallels transform to their anti-parallels
by a single unitary operator. 

To prove this we first show that for any three parallel states \( \left| \overrightarrow{n_{1}},\overrightarrow{n_{1}}\right\rangle  \),
\( \left| \overrightarrow{n_{2}},\overrightarrow{n_{2}}\right\rangle  \) and
\( \left| \overrightarrow{n_{3}},\overrightarrow{n_{3}}\right\rangle  \), the
Bloch vectors \( \overrightarrow{n_{1}} \), \( \overrightarrow{n_{2}} \) and
\( \overrightarrow{n_{3}} \) must lie on the same great circle if the parallel
states are to be transformed to their anti-parallels by the same unitary operator.
But before that note that any three (distinct) parallel states are linearly
independent. The reason is that all non-trivial linear combinations of any two
parallel states are entangled and hence no parallel state, which are product
states, lie in their linear span. Consider then any three parallel states \( \left| \overrightarrow{n_{i}},\overrightarrow{n_{i}}\right\rangle \: (i=1,2,3) \)
and suppose that there exists a unitary operator \( U^{\prime \prime } \) such
that
\begin{equation}
\label{4}
U^{\prime \prime }\left| \overrightarrow{n_{i}},\overrightarrow{n_{i}}\right\rangle \left| M\right\rangle =e^{i\vartheta _{i}}\left| \overrightarrow{n_{i}},-\overrightarrow{n_{i}}\right\rangle \left| N\right\rangle \: (i=1,2,3)
\end{equation}

The \( \overrightarrow{n_{i}} \)'s are constrained by the fact that \( U^{\prime \prime } \)
is unitary:
\begin{equation}
\label{5}
\left\langle \overrightarrow{n_{i}}\right. \left| \overrightarrow{n_{j}}\right\rangle =e^{i(\vartheta _{j}-\vartheta _{i})}\left\langle -\overrightarrow{n_{i}}\right. \left| -\overrightarrow{n_{j}}\right\rangle \: (i,j=1,2,3)
\end{equation}

where we have assumed that the \( \left| \overrightarrow{n_{i}}\right\rangle  \)'s
are mutually non-orthogonal. Taking 
\begin{equation}
\label{6}
\left| \overrightarrow{n_{i}}\right\rangle =e^{i\alpha _{i}}(\cos \frac{\theta _{i}}{2}\left| \overrightarrow{n_{1}}\right\rangle +e^{i\phi _{i}}\sin \frac{\theta _{i}}{2}\left| -\overrightarrow{n_{1}}\right\rangle )\: (i=2,3)
\end{equation}

and using (5), we have \( 2\alpha _{i}=\frac{1}{2}(\vartheta _{i}-\vartheta _{1})+(2k+1)\pi  \)
(with \( k \) an integer and \( i=2,3 \)) and 
\begin{equation}
\label{7}
\phi _{2}-\phi _{3}=n\pi \: (n=0,\pm 1,\pm 2,....)
\end{equation}

which means that \( \overrightarrow{n_{1}} \), \( \overrightarrow{n_{2}} \)
and \( \overrightarrow{n_{3}} \) lie on the same great circle.

In the special case when any two of \( \left| \overrightarrow{n_{1}}\right\rangle  \),
\( \left| \overrightarrow{n_{2}}\right\rangle  \) and \( \left| \overrightarrow{n_{3}}\right\rangle  \)
are orthogonal, the corresponding Bloch vectors always lie on a great circle
of the Bloch sphere and hence the corresponding parallel states can be transformed
to their anti-parallels by the unitary operator \( U^{\prime } \) of (2). Therefore
in any case, three parallel states can be transformed to their anti-parallels
by the same unitary operator if and only if the corresponding Bloch vectors
lie on a great circle of the Bloch sphere. The `if' part follows from the existence
of \( U^{\prime } \) of (2). 

An immediate consequence is that there cannot exist a measuring strategy on
an anti-parallel state chosen at random from \( A_{S_{L}} \) that estimates
\( \overrightarrow{n} \) better than any optimal strategy on a parallel state
chosen at random from \( P_{S_{L}} \). Being a little imprecise, more information
about \( \overrightarrow{n} \) is not extractable from \( A_{S_{L}} \) as
compared to \( P_{S_{L}} \). Gisin and Popescu had conjectured in {[}1{]} that
more information about \( \overrightarrow{n} \) \emph{is} extractable from
\( A_{T} \) as compared to \( P_{T} \) where \( T \) is the set of the four
vertices 
\begin{equation}
\label{8}
(0,0,1),\: (\frac{\sqrt{8}}{3},0,-\frac{1}{3}),\: (-\frac{\sqrt{2}}{3},\sqrt{\frac{2}{3}},-\frac{1}{3}),\: (-\frac{\sqrt{2}}{3},-\sqrt{\frac{2}{3}},-\frac{1}{3})
\end{equation}

on the Bloch sphere, of a tetrahedron. One can see that the elements of \( T \)
do not lie on any great circle whereby it is not possible to transform \( \left| \overrightarrow{n},\overrightarrow{n}\right\rangle  \)
to \( \left| \overrightarrow{n},-\overrightarrow{n}\right\rangle  \) for all
\( \overrightarrow{n}\in T \) by a single unitary operator. The conjecture
thus survives this attempt of defeat. Within a few paragraphs we provide more
basis for this conjecture. 

Before proceeding further, we parenthetically note that the largest set of \( \overrightarrow{n} \)'s
for which \( \left| \overrightarrow{n},-\overrightarrow{n}\right\rangle  \)
to \( \left| \overrightarrow{n},\overrightarrow{n}\right\rangle  \) is possible,
is again \( S_{L} \). The corresponding unitary operator is just the inverse
of the unitary operator \( U^{\prime } \) of (2). Such a transformation would
not be possible for a larger set of Bloch vectors as that would enable one to
transform a larger set of parallels to their anti-parallels (by the inverse
of that (unitary) transformation) which we have seen to be untrue. 

We have seen that \( S_{L} \) is the largest set of Bloch vectors for which
\( \left| \overrightarrow{n},\overrightarrow{n}\right\rangle  \) goes over
to \( \left| \overrightarrow{n},-\overrightarrow{n}\right\rangle  \) by the
same machine. We now want to see how far we can go by just flipping the second
state. Consider a unitary operator \( U_{f} \) such that

\[
U_{f}\left| 0\right\rangle \left| M\right\rangle =e^{i\vartheta _{0}}\left| 1\right\rangle \left| N\right\rangle \]
\begin{equation}
\label{9}
U_{f}\left| \overrightarrow{n}\right\rangle \left| M\right\rangle =e^{i\vartheta _{n}}\left| -\overrightarrow{n}\right\rangle \left| N\right\rangle 
\end{equation}

where 
\begin{equation}
\label{10}
\left| \overrightarrow{n}\right\rangle =e^{i\alpha }(\cos \frac{\theta }{2}\left| 0\right\rangle +e^{i\phi }\sin \frac{\theta }{2}\left| 1\right\rangle )
\end{equation}

Here again \( \left| N\right\rangle  \) is independent of the input qubits.

As \( U_{f} \) is unitary, we must have 
\begin{equation}
\label{11}
\left\langle 0\right. \left| \overrightarrow{n}\right\rangle =e^{i(\vartheta _{n}-\vartheta _{0})}\left\langle 1\right. \left| -\overrightarrow{n}\right\rangle 
\end{equation}

where we have assumed that \( \left\langle 0\right. \left| \overrightarrow{n}\right\rangle \neq 0 \).
This gives us 
\begin{equation}
\label{12}
2\alpha =\vartheta _{n}-\vartheta _{0}+(2r+1)\pi 
\end{equation}

with \( r \) an integer. This does not give any constraint on \( \overrightarrow{n} \)
which essentially depends on \( \theta  \) and \( \phi  \). If \( \left\langle 0\right. \left| \overrightarrow{n}\right\rangle =0 \),
then the orthogonal states \( \left| 0\right\rangle  \) and \( \left| \overrightarrow{n}\right\rangle  \)
can obviously be unitarily transformed to the orthogonal states \( \left| 1\right\rangle  \)
and \( \left| -\overrightarrow{n}\right\rangle  \). Thus in any case, there
exists a unitary operator to flip a qubit chosen at random from two arbitrary
but fixed qubits. Next let us assume that 
\begin{equation}
\label{13}
U_{f}\left| \overrightarrow{m}\right\rangle \left| M\right\rangle =e^{i\vartheta _{m}}\left| -\overrightarrow{m}\right\rangle \left| N\right\rangle 
\end{equation}

where
\begin{equation}
\label{14}
\left| \overrightarrow{m}\right\rangle =e^{i\alpha ^{\prime }}(\cos \frac{\theta ^{\prime }}{2}\left| 0\right\rangle +e^{i\phi ^{\prime }}\sin \frac{\theta ^{\prime }}{2}\left| 1\right\rangle )
\end{equation}

Unitarity restricts \( \overrightarrow{m} \) to lie on the same great circle
as of \( \overrightarrow{n} \) and the Bloch vector for \( \left| 0\right\rangle  \)
on the Bloch sphere. As \( \left| \overrightarrow{m}\right\rangle  \) can be
written as a linear combination of \( \left| 0\right\rangle  \) and \( \left| \overrightarrow{n}\right\rangle  \),
we must also check for linearity and constrain \( \overrightarrow{m} \) accordingly.
But surprisingly, as one can easily see, linearity does not give any new constraint.
Thus \( S_{L} \) is yet again the largest set of Bloch vectors for which the
corresponding \emph{qubits} can be flipped. Consequently, the unitary operators
for which \( \left| \overrightarrow{n},\overrightarrow{n}\right\rangle  \)
goes over to \( \left| \overrightarrow{n},-\overrightarrow{n}\right\rangle  \)
(and also \( \left| \overrightarrow{n},-\overrightarrow{n}\right\rangle  \)
to \( \left| \overrightarrow{n},\overrightarrow{n}\right\rangle  \)) for the
largest set of \( \overrightarrow{n} \)'s is surprisingly possible for the
form \( I\otimes U_{2} \) where \( I \) is the identity operator on the Hilbert
space of the first qubit and \( U_{2} \) is a unitary operator on the Hilbert
space of the second qubit. Thus we can actually transform \( \left| \overrightarrow{m},\overrightarrow{n}\right\rangle  \)
to \( \left| \overrightarrow{m},-\overrightarrow{n}\right\rangle  \) (and also
\( \left| \overrightarrow{m},-\overrightarrow{n}\right\rangle  \) to \( \left| \overrightarrow{m},\overrightarrow{n}\right\rangle  \))
for all \( \overrightarrow{n} \) lying on a great circle and for \emph{any}
\( \overrightarrow{m} \). 

As we have already transpired, we shall next consider the case of probabilistic
exact parallel to anti-parallel transformations. Two arbitrarily chosen parallel
states can always be unitarily transformed to the corresponding anti-parallel
states. This is because two arbitrarily chosen Bloch vectors always lie on a
great circle. The same is not true for \emph{three} arbitrarily chosen parallel
states. Hence it is relevant to consider whether a probabilistic (exact) transformation
is possible in this case. Precisely, we want to find out whether there exists
a unitary operator for which 
\begin{equation}
\label{15}
\left| \overrightarrow{n_{i}},\overrightarrow{n_{i}}\right\rangle \left| M\right\rangle \rightarrow \sqrt{\gamma _{i}}\left| \overrightarrow{n_{i}},-\overrightarrow{n_{i}}\right\rangle \left| M_{i}\right\rangle +\sqrt{1-\gamma _{i}}\left| \Phi _{i}\right\rangle 
\end{equation}

where \( \left| M\right\rangle  \), \( \left| M_{i}\right\rangle  \) are ancilla
states, \( \left| \Phi _{i}\right\rangle  \)'s belong to the combined Hilbert
space of the two qubits and ancilla, \( 0<\gamma _{i}<1 \) and \( I_{4}\otimes \left| M_{i}\right\rangle \left\langle M_{i}\right| \left| \Phi _{j}\right\rangle =0 \)
for \( i,j=1,2,3 \). A matrix theoretical argument following Duan and Guo {[}7,8{]}
shows that such an operator always exists and the corresponding optimal success
probabilities \( \gamma _{i} \) can also be found out. 

Let us now try to find whether there exists a unitary operator for which (15)
holds for \( i=1,2,...,k \) with \( k\geq 4 \). The corresponding parallel
states must be linearly dependent as all the parallel states of two qubits span
only a three dimensional subspace of the Hilbert space of two qubits. We assume
that the set \( S^{\prime } \) of \( \overrightarrow{n_{i}} \)'s for \( i=1,2,...,k \)
is not a subset of any great circle as in that case one can transform \( \left| \overrightarrow{n_{i}},\overrightarrow{n_{i}}\right\rangle  \)
to \( \left| \overrightarrow{n_{i}},-\overrightarrow{n_{i}}\right\rangle  \)
with unit probabilities. Now in a unitary-reduction process (15), the dimension
of the linear span of the domain must always be greater than or equal to the
dimension of the linear span of the range. Consequently there is no probabilistic
transformation taking the states of \( P_{S^{\prime }} \) to \( A_{S^{\prime }} \),
if the dimension of the linear span of \( A_{S^{\prime }} \) is four. 

Let \( S \) be any set of four or more Bloch vectors \( \overrightarrow{n} \)
which do not lie on any great circle and corresponding to which \( A_{S} \)
spans the whole Hilbert space of the two qubits. We conjecture that there exists
a measuring strategy on an anti-parallel state chosen at random from \( A_{S} \)
that estimates \( \overrightarrow{n} \) better than the optimal measuring strategy
on a parallel state chosen at random from \( P_{S} \). The result of Gisin
and Popescu {[}1{]}, for the universal case, gives support to this conjecture.
Again if \( S \) is any set of four Bloch vectors for which \( A_{S} \) is
a linearly independent set, there exists an optimal state discriminating strategy
{[}11{]} for \( A_{S} \) which give some kind of estimation of the states in
\( S \), while no such estimation is possible for \( P_{S} \) (as \( P_{S} \)
is a linearly dependent set) which provides further support to our conjecture.
The conjecture of Gisin and Popescu {[}1{]}, refered to earlier, is a special
case of this conjecture. This is because there are four Bloch vectors in \( T \)
and they do not lie on any great circle and the linear span of \( A_{T} \)
is of dimension four. 

For completeness, we note that linearly independent anti-parallel states can
always be probabilistically transformed to their parallels with non-zero success
probabilities (also non-unit, if the Bloch vectors do not lie on a great circle).

To summarize, we have shown that the Bloch vectors lying on a great circle of
the Bloch sphere is the largest set \( S_{L} \) for which the corresponding
parallels are exactly transformed to their anti-parallels. As a consequence,
with the apriori knowledge that an anti-parallel state belongs to \( A_{S_{L}} \),
one cannot estimate the Bloch vector better than the optimal case with the apriori
knowledge that a parallel state belongs to \( P_{S_{L}} \). We then found to
our surprise that the largest set of Bloch vectors for which the corresponding
qubits can be flipped by a single unitary operator is again \( S_{L} \). We
then showed that probabilistic exact parallel to anti-parallel transformation
is not possible if the corresponding anti-parallels span the whole four-dimensional
Hilbert space of the two qubits. This allowed us to provide ground for a conjecture
made by Gisin and Popescu {[}1{]} and also to make a more general conjecture.
To justify this general conjecture, we have to find out state estimation strategies
for finite sets of parallel and anti-parallel states.\\

The authors acknowledge useful discussions with Guruprasad Kar, Somshubhro Bandyopadhyay
and Debasis Sarkar. U.S. thanks Dipankar Home for encouragement and acknowledges
partial support by the Council of Scientific and Industrial Research, Government
of India, New Delhi.\\

\textbf{References:}

{[}1{]} N. Gisin and S. Popescu, Phys. Rev. Lett. \textbf{83} 432 (1999).

{[}2{]} S. Massar and S. Popescu, Phys. Rev. Lett. \textbf{74}, 1259 (1995).

{[}3{]} S. Massar, quant-ph/0004035.

{[}4{]} The normalized pure product state \( \left| \psi \right\rangle \otimes \left| \phi \right\rangle  \)
of two qubits, where the (unit) Bloch vectors of \( \left| \psi \right\rangle  \)
and \( \left| \phi \right\rangle  \) are \( \overrightarrow{n} \) and \( \overrightarrow{m} \)
respectively, is denoted here by \( \left| \overrightarrow{n},\overrightarrow{m}\right\rangle  \). 

{[}5{]} For any two non-orthogonal states \( \left| \overrightarrow{m}\right\rangle  \),
\( \left| \overrightarrow{n}\right\rangle  \), from (1) we get \( \left| \left\langle N(\overrightarrow{m})\right| \left. N(\overrightarrow{n})\right\rangle \right| =1 \).
Then using Cauchy's inequality we see that \( \left| N(\overrightarrow{m})\right\rangle  \)
and \( \left| N(\overrightarrow{n})\right\rangle  \) are equal upto a phase. 

{[}6{]} V. Buzek and M. Hillery, Phys. Rev. A, \textbf{54}, 1844 (1996); D.
Bruss et. al., Phys Rev. A, \textbf{57}, 2368 (1998).

{[}7{]} L. M. Duan and G. C. Guo, Phys. Lett. A, \textbf{243}, 261 (1998).

{[}8{]} L. M. Duan and G. C. Guo, Phys. Rev. Lett., \textbf{80}, 4999 (1998).

{[}9{]} W. K. Wootters and W. H. Zurek, Nature, \textbf{299}, 802 (1982).

{[}10{]} N. Gisin, private communication.

{[}11{]} A. Peres, Quantum Theory: Concepts and Methods, Kluwer 1993; A. Chefles
and S.M. Barnett, J. Mod. Opt., \textbf{45}, 1295 (1998).

\end{document}